\documentstyle[12pt]{article}
\newcommand{\A}{\alpha}
\newcommand{\ad}{{\rm ad}}

\newcommand{\bg}{{\bf g}}
\newcommand{\bG}{\bar{G}}
\newcommand{\cF}{{\cal F}}
\newcommand{\cM}{{\cal M}}
\newcommand{\CMP}{Commun. Math. Phys.}
\newcommand{\cN}{{\cal N}}
\newcommand{\CR}{\nonumber \\}

\newcommand{\D}{\delta}
\newcommand{\DE}{\Delta}
\newcommand{\E}{{\rm e}}
\newcommand{\EN}{\end{equation}}
\newcommand{\ENN}{\end{eqnarray}}
\newcommand{\ep}{\epsilon}
\newcommand{\EQ}{\begin{equation}}
\newcommand{\EQN}{\begin{eqnarray}}
\newcommand{\G}{\gamma}
\newcommand{\half}{{1\over2}}
\newcommand{\HBG}{\hat{{\bf g}}}
\newcommand{\I}{{\rm i}}
\newcommand{\IJMPA}[1]{Int. J. Mod. Phys. {\bf A#1}}
\newcommand{\mat}[2]{\left(\begin{array}{#1}#2\end{array}\right)}
\newcommand{\MPLA}[1]{Mod. Phys. Lett. {\bf A#1}}
\newcommand{\NPB}[1]{Nucl. Phys. {\bf B#1}}
\newcommand{\pa}{\partial}
\newcommand{\PL}{Phys. Lett.}
\newcommand{\PLB}[1]{Phys. Lett. {\bf B#1}}
\newcommand{\Ra}{{ }^{r}a}

\newcommand{\Rchi}{{ }^{r}\chi}

\newcommand{\RG}{{ }^{r}G}
\newcommand{\Rxi}{{ }^{r}\xi}
\newcommand{\str}{{\rm str}}

\newcommand{\Tb}{{ }^{t}b}
\newcommand{\TbG}{{ }^{t}\bG}
\newcommand{\Tbchi}{{ }^{t}\bar{\chi}}
\newcommand{\Teta}{{ }^{t}\eta}

\newcommand{\tr}{{\rm tr}}
\newcommand{\TT}{{\tilde T}}

\newcommand{\LM}{\Lambda}

\newcommand{\VP}{\varphi}
\relax
\begin{document}
\renewcommand{\thefootnote}{\fnsymbol{footnote}}
\begin{titlepage}
\begin{flushright}
NBI-HE-92-12  \\
February, 1992
\end{flushright}
\vskip 2em
\begin{center}
{\LARGE
Hamiltonian Reduction and Classical  \\
Extended Superconformal Algebras
\par}
\vskip 3em
\lineskip .75em
\normalsize
\begin{tabular}[t]{c}
{\large Katsushi Ito}
and
{\large Jens Ole Madsen} \\ \\
{\sl The Niels Bohr Institute} \\
{\sl University of Copenhagen} \\
{\sl Blegdamsvej 17, DK-2100} \\
{\sl Copenhagen {\O}, Denmark}
\end{tabular}
\vskip 1.5em
{\bf Abstract}
\end{center} \par
We present a systematic construction of classical extended
superconformal algebras from the hamiltonian reduction of a class of
affine Lie superalgebras, which include an even subalgebra
$sl(2)$.
In particular, we obtain the doubly extended $N=4$ superconformal
algebra $\tilde{A}_{\G}$ from the hamiltonian reduction of the
exceptional Lie superalgebra $D(2|1;\G/(1-\G))$.
We also find the Miura transformation for these algebras and
give the free field representation.
A $W$-algebraic generalization  is discussed.
\end{titlepage}
\renewcommand{\thefootnote}{\arabic{footnote}}
\setcounter{footnote}{0}
\baselineskip=0.8cm
In the last few years it has been known that various higher-spin
extension of the Virasoro algebra including Zamolodchikov's $W$
algebra\cite{Za} can be constructed using the Drinfeld-Sokolov
type hamiltonian reduction of affine Lie algebras
(or the associated integrable systems) \cite{DrSo},\cite{Be} and
its generalization \cite{Po}-\cite{BaTjDr}.
Supersymmetric extension of the Virasoro algebra is important
from the viewpoint of application to superstrings and topological
field theory.
It has been shown that $N=1$ and $N=2$ super $W$ algebras can be
obtained from the hamiltonian reduction (or related super Toda
field theories) of a special class of affine Lie superalgebras
which have the purely odd simple root system \cite{EvHo}.

The purpose of this paper is to present a systematic construction
of the classical extended superconformal algebra (ESA) from the
hamiltonian reduction viewpoint.
ESAs with $u(N)$ and $so(N)$ Kac-Moody symmetries, which  have been
constructed by Knizhnik and Bershadsky \cite{KnBe}, have in general
non-linear properties similar to those of the $W$-algebra.
Some previous works describe the relation between $N=1$ and 2
superconformal algebras and Lie superalgebras $osp(N|2)$ \cite{Kup}.
In particular Mathieu has shown that the hamiltonian structure
of the $osp(N|2)$ KdV equations leads to the classical $so(N)$
ESA \cite{Mat}, although in his formulation the Lie superalgebraic
structure is not obvious.

In the present paper we shall discuss a class of Lie superalgebras
which include even subalgebras $\bg\oplus sl(2)$, where
$\bg$ is a semi-simple Lie algebra.
Using Kac's notation \cite{Kac}, such basic classical Lie
superalgebras are classified as follows (see Table 1):
$A(n|1)$ $(n\geq 1)$, $A(1|0)$ $B(n|1)$, $D(n|1)$,
$B(1|n)$, $D(2|1;\A)$, $F(4)$ and $G(3)$.
By considering the hamiltonian reduction of an even subalgebra
$sl(2)$, we will find ESAs with
$A_{n}\oplus u(1)$ ($A_{1}$ for $n=1$), $u(1)$, $B_{n}$, $D_{n}$,
$C_{n}$, $A_{1}\oplus A_{1}$, $spin(7)$, $G_{2}$ Kac-Moody symmetry
\footnote{For $B(1|n)$ the corresponding algebra has
spin 0, 1 and 2 fermionic currents with $C_{n}$ Kac-Moody symmetry.}.
In particular for $D(2|1;\A)$ the corresponding ESA is shown to be the
$N=4$ doubly extended superconformal algebra \cite{GoSc}.
We will also show that the Miura transformation can be naturally
obtained by connecting the Drinfeld-Sokolov gauge and the
\lq\lq diagonal" gauge.
The present classical formulation is a crucial step for the
study of the quantum ESA using the Coulomb gas representation.
We note that a bosonic analogue of $u(N)$ ESA has been discussed
in refs. \cite{BaTjDr} by considering the hamiltonian reduction
of $sl(N+2)$.
However, for other Lie superalgebras there is no bosonic counterpart,
especially for $D(2|1;\A)$.

We begin by explaining the hamiltonian reduction by taking an affine Lie
superalgebra $\HBG=A(n|1)^{(1)}$ at level $k$
associated with a basic classical Lie superalgebra
$\bg=A(n|1)=sl(N|2)$ $(N=n+1)$ as an example.
The algebra $\HBG$ consists of an element $(X(z), x_{0})$  of the form
\EQ
X(z)=\mat{c|c}{X_{1}(z) & \xi(z) \\
                  \hline
                  \eta(z) & X_{2}(z) }, \quad \quad
\str X=\tr X_{1}-\tr X_{2} =0,
\label{eq:sln2}
\EN
where $X_{1}$ is a $N\times N$ matrix and $X_{2}$ is a
$2\times 2$ matrix, whose components are commuting numbers.
$\xi=(\xi_{1},\xi_{2})$ and
$\eta=\mat{c}{\Teta_{1} \\ \Teta_{2}}$
are $N\times 2$ and $2\times N$ matrices and
$\xi_{i}$ and $\eta_{i}$ are fermionic $N$-vectors.
$x_{0}$ is a number.
The commutation relation is defined by
\EQ
\mbox{[} (X,x_{0}), (Y, y_{0}) \mbox{]}
=( \mbox{[} X, Y \mbox{]} ,
   k \int {d z \over 2\pi \I } \str (X \pa Y) ).
\EN
The dual space $\HBG^{*}$ of the superalgebra $\HBG$
consists of an element $(J(z),\A_{0})$, where the scalar product
$\langle \ ,\  \rangle$ take the form
\EQ
\langle (J, \A_{0}), (X,x_{0})\rangle
=\int {d z \over 2\pi \I } \str (J X)+\A_{0} x_{0}.
\EN
Let us parametrize $J(z)$ as
\EQ
J(z)=\mat{c|c}{ A(z) & a(z) \\  \hline
                b(z) & B(z) }, \quad \quad
{\rm str}J=\tr A-\tr B =0.
\EN
where  $A(z)\in gl(N)$, $B(z)\in gl(2)$,
$a=(a_{1}(z), a_{2}(z))$,
$b=\mat{c}{ \Tb_{1}(z) \\ \Tb_{2}(z)}$.
$a_{i}$ and $b_{i}$ $(i=1,2)$ are $N$-dimensional vectors whose
components are fermionic currents.
$\HBG$ acts on $\HBG^{*}$ as an coadjoint action $\ad^{*}$
(or infinitesimal gauge transformation $\D$) defined by the
formula
\EQ
\langle\ad^{*}(X,x_{0})(J, \A_{0}), (Y,y_{0})\rangle
=-\langle (J,\A_{0}), \mbox{[}(X,x_{0}), (Y,y_{0}) \mbox{]}\rangle ,
\EN
\EQ
i.e. \quad \quad \quad
\ad^{*}(X,x_{0})(J,\A_{0})=(\mbox{[}X, J\mbox{]}+\A_{0} k\pa X, 0).
\label{eq:coa}
\EN
{}From (\ref{eq:sln2}) and (\ref{eq:coa}),
the gauge transformation $\D J=[ X, J ]+\A_{0} k\pa X $ becomes
\EQ
\D J=\mat{c|c}{
[ X_{1},A ]-a\eta+\xi b +\A_{0}k \pa X_{1} &
-A \xi+\xi B -a X_{2}+X_{1}a +\A_{0}k \pa\xi \\ \hline
-B \eta+\eta A -b X_{1}+X_{2}b +\A_{0}k \pa\eta &
[ X_{2},B ]-b \xi+\eta a +\A_{0}k \pa X_{2} }.
\label{eq:gaug}
\EN
Next we impose the constraint on $\HBG^{*}$.
In the present paper, we shall consider the constraint for
the $sl(2)$ even subalgebra of $\bg$ and keep the other
even affine algebra symmetry.
We may choose the following constraint set $\cM$
(or the set of levels), which consists of the elements of the form
$(J(z), 1)$
\EQ
J(z)=\mat{c|c}{ A(z) & a(z) \\ \hline
                b(z) & B(z) }, \quad
B(z)=\mat{cc}{ * & 1 \\
                {*} & *}.
\label{eq:const}
\EN
Let $\cN$ be the space of gauge transformations which preserve this
constraint.
The space $\cN$ is determined by the condition $\D B_{12}=0$
for arbitrary $J(z) \in \cM$.
{}From the formula (\ref{eq:gaug}),
we get $(X_{2})_{12}=0$ and $\xi_{2}=0$ and $\eta_{1}=0$.
Namely $\cN$ can be expressed as $\cN=\cN_{u(N)}\cN_{0}$, where
$\cN_{u(N)}$ is the $u(N)$ gauge transformation group and $\cN_{0}$
consists of the elements $\E^{X}$ with
\EQ
X=\mat{c|cc}{0 & * & 0 \\
                \hline
                0 & 0 & 0 \\
                {*} & * & 0}.
\label{eq:gapa}
\EN
Let us define the reduced phase space by $\cF=\cM/\cN_{0}$.
We can take the following two typical gauges.
One is the \lq\lq Drinfeld-Sokolov" gauge:
\EQ
J_{DS}(z)=\mat{c|cc}{ W & G & 0 \\
                   \hline
                    0 & 0 & 1 \\
                   \TbG & T & U },
\quad W=J+{U \over N} 1_{N},
\EN
where $J$ denotes the $sl(N)$ Kac-Moody current at level $k$ and
$U$ is the $u(1)$ current, $1_{N}$ is the $N\times N$ unit matrix.
The currents $W$ generate the $u(N)$ Kac-Moody algebra.
The other is the \lq\lq diagonal" gauge
\EQ
J_{diag}(z)=\mat{c|cc}{ A  & 0     & \chi  \\ \hline
                    \Tbchi & p_{1} & 1     \\
                        0  & 0     & p_{2} },
\quad A=\hat{J}+{p_{1}+p_{2} \over N} 1_{N}.
\EN
Here $\hat{J}$ is the $sl(N)$ current at level $\hat{k}$.
A Poisson bracket structure on the reduced phase space $\cF$ is
determined by the coadjoint action preserving the gauge.
This method is known Polyakov's \lq\lq soldering"
procedure \cite{Po}.
To do this, it is convenient to  decompose the gauge parameter
by taking
$X=\LM_{1}+\LM_{2}+\LM_{3}$ where
\EQ
\LM_{1}=\mat{c|c}{\ep_{1} & 0 \\ \hline
                        0 & 0  } ,
\quad
\LM_{2}=\ep \mat{c|c}{1_{N} & 0 \\ \hline
                         0  & {N\over 2}1_{2}} ,
\quad
\LM_{3}=\mat{c|c}{0  & \xi \\ \hline
                \eta & \ep_{2}  },
\label{eq:para}
\EN
and $\ep_{1}\in sl(N)$,
$\ep_{2}=\mat{cc}{\ep_{11} & \ep_{12} \\
                  \ep_{21} & -\ep_{11}} \in sl(2)$.
$\LM_{1}$ and $\LM_{2}$ correspond to the
$sl(N)$ Kac-Moody symmetry and the $u(1)$ symmetry, respectively.
The parameter $\LM_{3}$ belongs to the $sl(2)$ and fermionic part.
{}From the conditions preserving the Drinfeld-Sokolov gauge:
\EQ
\D a_{2}=\D \Tb_{1}=0, \ \ \ \
\D B_{11}=\D B_{12}=0,
\EN
we can express the gauge parameters
$\xi_{1}$, $\eta_{2}$, $\ep_{11}$ and $\ep_{21}$ in terms of
$\xi_{2}$, $\eta_{1}$, $\ep$ and $\ep_{12}$.
Therefore the gauge transformations on the reduced phase space
can be expressed by using these parameters:
\EQN
\D U &=& N k\pa\ep +\Teta_{1}G -\TbG\xi_{2},             \CR
\D J &=& [ \ep_{1}, J ] +k\pa\ep_{1}
         -G\Teta_{1} +\xi_{2}\TbG
         -{1\over N} (\Teta_{1}G-\TbG \xi_{2})1_{N}      \CR
\D G &=& -k^{2}\pa^{2}\xi_{2} +k(2W-U)\pa\xi_{2}
         +\left\{ W(U-W) +k\pa (W-U) +T\right\} \xi_{2}  \CR
     & & +{3k \over 2} G \pa \ep_{12}
         +\left\{ ({1\over2}U-W) G +k \pa G \right\} \ep_{12}
         -({N\over 2}-1)\ep G +\ep_{1} G,                \CR
\D \TbG
    &=&  k^{2}\pa^{2} \Teta_{1}
         +k\pa\Teta_{1} (2W-U)
         +\Teta_{1}\left\{ -T+W(W-U)+k \pa W\right\}     \CR
    & &  +{3k\over 2} \TbG \pa \ep_{12}
         +\left\{ \TbG (W-{U\over 2})+k\pa\TbG\right\} \ep_{12}
         -\TbG\ep_{1} +({N\over 2}-1)\TbG\ep,            \CR
\D T&=&  {N\over 2}k^{2}\pa^{2}\ep -{N\over 2}k U\pa\ep  \CR
    & &  -{k^{3}\over 2}\pa^{3}\ep_{12}
         +\left( 2k T +{k\over 2}U^{2}-k^{2}\pa U\right)\pa\ep_{12}
         +\left( k\pa T +{k\over 2}U\pa U -{k^{2}\over 2}\pa^{2} U
         \right) \ep_{12}                                \CR
    & &  +\Teta_{1}\left\{(W-U)G+k\pa G\right\}+2k\pa\Teta_{1} G
         +\TbG (U-W) \xi_{2} +k \TbG \pa \xi_{2}.
\label{eq:tra}
\ENN
Note that $T$ itself does not behave as energy-momentum tensor.
In order to get the correct form,
we must deform $T$ by a differential polynomial of $U$, so we
define $\TT=T+{1\over 4} U^{2}-{k \over 2}\pa U$.
{}From (\ref{eq:tra}) we can show that $\TT$ behaves as
\EQN
\D \TT
&=&\Teta_{1}\left\{(W-{U\over 2})G+{k\over 2}\pa G\right\}
     +{3k\over 2}\pa \Teta_{1} G
     +\left\{ \TbG ({U\over 2}-W)+{k\over 2}\pa \TbG\right\}\xi_{2}
     +{3k\over 2} \TbG \pa \xi_{2}          \CR
& &  -{k^{3} \over 2} \pa^{3} \ep_{12}
     +2k \TT \pa \ep_{12}
     +k \pa \TT \ep_{12}.
\ENN
By expressing the infinitesimal gauge transformation $\D$ as
\EQ
\D= \int {d z \over 2 \pi \I}
    \str
    \mat{c|cc}{W & G & 0 \\ \hline
                  0 & 0 & 1 \\
                \TbG & \TT & U }
    \mat{c|cc}{\ep_{1} +\ep 1_{N}& 0 & \xi_{2} \\ \hline
             \Teta_{1} & {N\over 2}\ep  & \ep_{12} \\
             0  & 0  & {N\over 2}\ep },
\EN
we find the operator product expansions for $U,J,G,\bG$ and $\TT$,
\EQN
\TT(z)\TT(w)
&=& {3k^{3} \over(z-w)^{4}}
   +{-2 k \TT  \over (z-w)^{2}}
   +{-k \pa \TT \over z-w}+\cdots ,                \CR
\TT(z)G_{i}(w)
&=& {{-3k\over 2}G_{i}(w) \over (z-w)^{2}}
   +{ J_{i j}G_{j} +({1\over N}-\half )U G_{i}
     -k\pa G_{i} \over z-w}+\cdots,                \CR
\TT(z)\bG_{i}(w)
&=& {{-3k\over 2}\bG_{i}(w) \over (z-w)^{2}}
   +{ \bG_{j}J_{j i} -({1\over N}-\half )U\bG_{i}
      -k\pa \bG_{i} \over z-w}+\cdots ,            \CR
G_{i}(z)\bG_{j}(w)
&=& { -2 k^{2}\D_{i j} \over (z-w)^{3}}
   +{k(U\D_{i j}-2W_{i j})\over (z-w)^{2}}
   +{T\D_{i j}-W_{i k}W_{k j}+W_{i j} U -k\pa W_{i j}
      \over z-w}+\cdots ,                          \CR
J_{k l}(z)G_{i}(w)
&=&
{\D_{i l} G_{k} -{1\over N}G_{i}\D_{k l}\over z-w}+\cdots, \quad
J_{k l}(z)\bG_{i}(w)
={-\D_{i k} \bG_{l} +{1\over N}\bG_{i}\D_{k l}\over z-w}+\cdots ,
\CR
J_{ij}(z)J_{kl}(w)
&=&{k(\D_{i l}\D_{j k}-{1\over N} \D_{i j}\D_{k l})\over (z-w)^{2}}
  +{\D_{j k}J_{i l}-\D_{i l}J_{k j} \over z-w}+\cdots,  \quad
U(z)U(w)={{-2N k \over N-2 }\over (z-w)^{2}}+\cdots,  \CR
U(z)G_{i}(w)&=&{-G_{i}(w) \over z-w}+\cdots, \quad  \quad
U(z)\bG_{i}(w)={\bG_{i}(w) \over z-w}+\cdots .
\ENN
The total energy-momentum tensor $T_{total}$ is defined by
adding the Sugawara form of energy-momentum tensors of the $sl(N)$
and $u(1)$ Kac-Moody algebras;
\EQ
T_{total}(z)={1\over -k}\TT(z)+T_{sl(N)}(z)+T_{u(1)}(z),
\label{eq:em}
\EN
where $T_{sl(N)}={1\over 2k} {\rm tr}J^{2}$
and $T_{u(1)}=-{N-2 \over 4N k}U^{2}$.
The supercurrents $G_{i}(z)$ and $\bG_{i}(z)$ have conformal
weight 3/2 with respect to $T_{total}(z)$.
These operator product expansions are nothing but those of the
classical $u(N)$ extended superconformal algebra.
In the case of $N=2$ the $u(1)$ current decouples
from the algebra and we get the $N=4$ $sl(2)$ ESA.

Next we discuss the Miura transformation by considering the system of
linear differential equations \cite{Be}:
\EQ
(k \pa -J(z) )v(z) =0,  \quad
v(z)=\mat{c}{ v_{0}  \\ \hline v_{1} \\ v_{2} },
\EN
where $J\in \cM$ and $v_{0}$ is a $N$-vector.
Under the gauge transformation
\EQ
J(z)\rightarrow n(z) J(z) n(z)^{-1} +k  (\pa n(z))n(z)^{-1},
\quad v \rightarrow n(z)v, \quad n(z) \in \cN_{0},
\EN
the component $v_{1}$ is gauge invariant.
Therefore we can write the system of linear differential equations
into the gauge-invariant higher (pseudo-)differential
operator acting on the component $v_{1}$.
In the Drinfeld-Sokolov gauge, the differential equations becomes
\EQ
k \pa v_{0}=W v_{0}+G v_{1}, \quad
k \pa v_{1}= v_{2}, \quad
k \pa v_{2}= \TbG v_{0}+T v_{1} +U v_{2}.
\EN
By solving these equations in terms of $v_{1}$ we get
\EQ
\left\{ (k \pa)^{2} - U k \pa -T -\TbG (k \pa -W)^{-1} G \right\}
v_{1}=0.
\label{eq:DSLax}
\EN
Here we define the inverse of  $(k \pa -W)$ by
\EQ
(k \pa -W)^{-1}=(k\pa)^{-1}
\left\{1+\sum_{i=1}^{\infty}\left(W (k\pa)^{-1} \right)^{i} \right\} .
\EN
In the  \lq\lq diagonal" gauge, from the equations
\EQ
k \pa v_{0}= A  v_{0}+\chi v_{2}, \quad
k \pa v_{1}= \Tbchi v_{0} +p_{1} v_{1}+v_{2}, \quad
k \pa v_{2}= p_{2} v_{2},
\EN
we get
\EQ
\left\{ (k \pa-p_{2})(k \pa-p_{1})
  - (k\pa-p_{2})\Tbchi (k \pa-A +\chi \Tbchi)^{-1} \chi
(k \pa-p_{1})  \right\} v_{1}=0.
\label{eq:DiLax}
\EN
Comparing these two equations (\ref{eq:DSLax}) and (\ref{eq:DiLax}),
we get the free field representation of the $u(N)$ extended
superconformal algebra:
\EQN
U&=&p_{1}+p_{2}+\Tbchi \chi,
\quad \quad
J=\hat{J}-\chi \Tbchi -{\Tbchi \chi \over N}1_{N}, \CR
T&=&-p_{1}p_{2}+k \pa p_{1} +(p_{1}+p_{2})\Tbchi \chi
   -k (\pa \Tbchi)\chi -\Tbchi W \chi , \CR
G&=&p_{1}\chi+k \pa \chi-W\chi, \quad
\TbG=p_{2}\Tbchi-k \pa \Tbchi-\Tbchi W .
\ENN
The level $\hat{k}$ of the $sl(N)$ Kac-Moody current $\hat{J}$
is equal to $k-1$.
{}From eq. (\ref{eq:em}) the total energy momentum tensor becomes
\EQ
T_{total}=-\half (\pa \VP)^{2}-{1\over2}(\pa\phi)^{2}
           -\sqrt{k\over 2}\pa^{2}\VP
            +\half
              \left( (\pa \Tbchi) \chi - \Tbchi \pa \chi \right)
             +{1\over 2k}{\rm tr} J^{2},
\EN
where
$p_{1}-p_{2}=\sqrt{2k}\pa\VP$ and
$p_{1}+p_{2}=\sqrt{{2N k \over N-2}}\pa \phi$.

In a similar way, we may construct the classical $so(N)$ extended
superconformal algebra from the affine Lie superalgebra
$osp(N|2)^{(1)}$.
An element of $\bg=osp(N|2)$ is given by \cite{Kac}
\EQ
X=\mat{c|cc}{ A & a \\ \hline
                 b & B },
\quad
a=(a_{1}, a_{2}),  \quad
b=\mat{c}{ -\Ra_{2} \\ \Ra_{1}},
\EN
where $A\in so(N)$, $B\in C_{1}=A_{1}$ and
$a_{i}$ $(i=1,2)$ are $N$-dimensional fermionic vectors.
Here we define a $m\times n$ matrix ${}^{r}A$  of a $n\times m$
matrix $A=(a_{i,j})$ by ${}^{r}A=(a_{m+1-j, n+1-i})$.
We consider the same type of constraints $\cM$ as (\ref{eq:const}).
The gauge group $\cN$ which preserve the constraints $\cM$ is generated
by (\ref{eq:gapa}).
The soldering procedure can be done by replacing $\TbG$
by $\RG$, $\Tbchi$ by $-\Rchi$ and $p_{1}$ ($p_{2}$)
by $p$ ($-p$)
in the $sl(N|2)$ case,
we show only the results
\EQN
\D W&=&[ \ep_{1},W ] +k \pa\ep_{1}+G\Rxi_{2} +\xi_{2}\RG \CR
\D G&=& -k^{2} \pa^{2}\xi_{2} +k W \pa\xi_{2}
       +\left\{ T-W W +k \pa W \right\} \xi_{2}
      +{3k \over 2} G \pa \ep_{12}
      +\left( k \pa G - W G\right) \ep_{12} +\ep_{1} G \CR
\D T&=&{-k^{3} \over 2} \pa^{3} \ep_{12}
      +2k T \pa \ep_{12} +  k \pa T \ep_{12}
      -\Rxi_{2}(k\pa G+2W G) -3k\pa \Rxi_{2} G.
\ENN
Here we parametrize the gauge parameters by $\LM=\LM_{1}+\LM_{3}$
in (\ref{eq:para}) and put $\eta$ equal to
$\mat{c}{-\Rxi_{2} \\ \Rxi_{1}}$.
The Miura transformation
\footnote{
We note that the Miura transformation proposed in refs. \cite{ChKu}
($sl(2|2)$ case) and \cite{Mat} ($so(N)$ case) is apparently different
from the present one.
However if we define the path-ordered  exponential \lq\lq Pexp"
satisfying
\EQ
\pa {\rm P exp}( \int^{z} W(z')d z' ) =
{\rm P exp}( \int^{z} W(z')d z' )  W(z),
\EN
both Miura transformations are shown to be equivalent by virtue of the
formula
\EQ
(k \pa -W)^{-1}={\rm P exp}(-\int^{z} W(z')d z' )\pa^{-1}
                {\rm P exp}( \int^{z} W(z')d z' ).
\EN }
is given by
\EQ
(k \pa)^{2}-T -\RG (k \pa -W)^{-1} G
  =(k \pa+p)(k \pa-p)
  - (k\pa+p)\Tbchi (k \pa-A +\chi \Rchi)^{-1} \chi (k \pa-p).
\EN
Therefore the free field representation of the classical
$so(N)$ extended superconformal algebras is expressed as
\EQN
W&=& \hat{J}-\chi \Rchi, \quad
G= \sqrt{{k\over 2}}\pa \VP \chi +k \pa \chi -W \chi, \CR
T_{total}&=& -\half (\pa \VP)^{2}-\sqrt{{k\over 2}}\pa^{2}\VP
             -\Rchi \pa \chi +{1\over 2k}\tr J^{2},
\ENN
where  $\hat{J}$ is the $so(N)$ current in the diagonal gauge
and the total energy-momentum tensor $T_{total}$ is defined by
$-T/k +\tr W^{2} /(2k)$.

We proceed to the exceptional Lie superalgebra $D(2|1;\A)$.
The algebra contains even subalgebras
$A_{1}\oplus A_{1} \oplus A_{1}$ and has rank three.
Let $\A_{i}$ ($i=1,2,3$) be simple roots of $D(2|1;\A)$ with
the inner products
$\A_{1}^{2}=0$,  $\A_{2}^{2}=-2\G$, $\quad \A_{3}^{2}=-2(1-\G)$,
$\A_{1}\cdot\A_{2}=\G$, $\A_{1}\cdot\A_{3}=1-\G$
and $\A_{2}\cdot\A_{3}=0$.
By constructing the Cartan matrix, we find
that the parameter $\A$ is given by the relation $\A=\G /(1-\G)$.
The positive even roots are $\A_{2}$, $\A_{3}$ and
$2\A_{1}+\A_{2}+\A_{3}$, which generate $A_{1}$ even subalgebras.
The odd positive roots are $\A_{1}$, $\A_{1}+\A_{2}$, $\A_{1}+\A_{3}$
and $\A_{1}+\A_{2}+\A_{3}$.
For $\A=1$ the hamiltonian reduction can be done by using an explicit
$6\times 6$ matrix representation because $D(2|1;\A=1)=osp(4|2)$.
But for generic $\A$, there is no such representation.
Hence we express the Kac-Moody current as
\EQ
J(z)=\sum_{\A \in \DE_{\bar{0}}}J_{\A}e_{\A}
     +\sum_{\A \in \DE_{\bar{1}}}j_{\A}e_{\A}
     +\sum_{i=1}^{3} H^{i}h^{i},
\EN
and  a gauge parameter $X(z)$ as
\EQ
X(z)=\sum_{\A \in \DE_{\bar{0}}}\ep_{\A}e_{\A}
     +\sum_{\A \in \DE_{\bar{1}}}\xi_{\A}e_{\A}
     +\sum_{i=1}^{3} \ep^{i}h^{i}.
\EN
Here $\{e_{\A} , h^{i} \}$ are the Cartan-Weyl basis and
$\DE_{\bar{0}}$ ($\DE_{\bar{1}}$) is the set of even (odd) roots.
We denote the currents $J_{\pm\A_{2}}$, $\A_{1}\cdot H$
as $J_{\pm 2}$, $H_{1}$, etc. for simplicity.
Now we consider the constraint on an $A_{1}$ subalgebra
$\{ J_{1123},J_{-1123}, H_{1123} \}$ by imposing the condition
$J_{1123}(z)=1$.
On the reduced phase space we may take the Drinfeld Sokolov gauge
\EQN
J_{-1123}(z)&=&T(z), \quad
H_{1123}(z)=0, \quad
j_{-1}(z)=G_{1}(z), \CR
j_{-12}(z)&=&G_{12}(z), \quad
j_{-13}(z)=G_{13}(z), \quad
j_{-123}(z)=G_{123}(z),
\ENN
By using the structure constants of the algebra,
we can calculate the gauge transformation on the reduced phase space.
For Kac-Moody currents the results are
\EQN
\D H_{2}&=&2\G J_{2}\ep_{-2}-2\G J_{-2}\ep_{2}
                   +k\pa (\A_{2} \cdot \ep)
                   - \G G_{1}\xi_{1} +\G G_{12}\xi_{12}
                   -\G G_{13}\xi_{13}+\G G_{123}\xi_{123}, \CR
\D J_{2}&=&\G G_{13} \xi_{123}-\G G_{1}\xi_{12}
          -H_{2} \ep_{2}+k \pa \ep_{2}
          +(\A_{2}\cdot \ep)J_{2}, \CR
\D J_{-2}&=&\G G_{123} \xi_{13}- \G G_{12}\xi_{1}
          +H_{2} \ep_{2}+k \pa \ep_{-2}
          -(\A_{2}\cdot \ep)J_{-2}.
\ENN
The transformations of the currents $J_{\pm3}$ and $\A_{3}\cdot H$
can be obtained by replacing indices 2 (3) by 3 (2) and $\G$ by $1-\G$.
The supercurrents and the energy-momentum tensor transform  as
\EQN
\D G_{1}&=& J_{2}\xi_{-12}+J_{3}\xi_{-13}-T\xi_{123}
            +k \pa \xi_{-1}-\half (H_{2}+H_{3}) \xi_{-1} \CR
        & & -G_{12}\ep_{2}
            -G_{13}\ep_{3}
            -(\A_{1}\cdot \ep) G_{1}, \CR
\D G_{12}&=&-J_{3}\xi_{-123} +J_{-2}\xi_{-1}-T \xi_{13}
          +k \pa \xi_{-12}
          +\half (H_{2}-H_{3}) \xi_{-12}   \CR
         & & +G_{123}\ep_{3}
          -G_{1}\ep_{-2}
          -(\A_{12}\cdot \ep) G_{12}, \CR
\D G_{13}&=&-J_{2}\xi_{-123}+J_{-3}\xi_{-1} -T \xi_{12}
          +k \pa \xi_{-13}
          -\half (H_{2}-H_{3})\xi_{-13} \CR
         & & +G_{123}\ep_{2}
          -G_{1}\ep_{-3}
          -(\A_{13}\cdot \ep) G_{13}, \CR
\D G_{123}&=&-J_{-3}\xi_{-12} -J_{-2}\xi_{-13}-T \xi_{1}
          +k \pa \xi_{-123}
          +\half (H_{2}+H_{3}) \xi_{-123}  \CR
          & &+G_{13}\ep_{-2}
          +G_{12}\ep_{-3}
          -(\A_{123}\cdot \ep) G_{123}, \CR
\D T &=& -(\A_{1123}\cdot \ep)T+k\pa \ep_{-1123}
       +G_{123}\xi_{-1}
       +G_{13}\xi_{-12}
       +G_{12}\xi_{-13}
       +G_{1}\xi_{-123},
\ENN
where $\A_{1123}\cdot \ep=-k \pa \ep_{1123}$ and
\EQN
\xi_{-123} &=&-J_{-2}\xi_{12}-J_{-3}\xi_{13}+G_{123}\ep_{1123}
            +k \pa \xi_{1} +\half (H_{2}+H_{3})\xi_{1}, \CR
\xi_{-13} &=&-J_{2}\xi_{1}+J_{-3}\xi_{123}+G_{13}\ep_{1123}
            +k \pa \xi_{12} -\half (H_{2}-H_{3})\xi_{12}, \CR
\xi_{-12} &=&-J_{3}\xi_{1}+J_{-2}\xi_{123}+G_{12}\ep_{1123}
            +k \pa \xi_{13} -\half (-H_{2}+H_{3})\xi_{13}, \CR
\xi_{-1} &=&J_{2}\xi_{13}+J_{3}\xi_{12}+G_{1}\ep_{1123}
            +k \pa \xi_{123} -\half (H_{2}+H_{3})\xi_{123}, \CR
\ep_{-1123}&=& \ep_{1123} T
-\half\left( G_{1}\xi_{1}+G_{12}\xi_{12}+G_{13}\xi_{13}
+G_{123}\xi_{123}\right)
-\half k^{2} \pa^{2} \ep_{1123}.
\ENN
These relations are those of the non-linear $N=4$ doubly ESA
$\tilde{A}_{\G}$ \cite{GoSc} \cite{PeTa} obtained  by factorizing
fermions from the linear $N=4$ ESA $A_{\G}$ \cite{SeTrPr}.
Generators of $\tilde{A}_{\G}$  are two $sl(2)$ affine Lie algebras
$\{ J_{2}, J_{-2},  H_{2} \}$ at level $k\G$,
$\{ J_{3}, J_{-3},  H_{3} \}$ at level $k(1-\G)$,
four supercurrents $G_{1}$, $G_{12}$, $G_{13}$ and $G_{123}$
and the total energy-momentum tensor $T_{total}=-T/k +T_{Sugawara}$.

One can also calculate the extended superconformal algebras
with $B_{3}$ and $G_{2}$ Kac-Moody symmetries from the hamiltonian
reduction of affine Lie superalgebras associated with exceptional
Lie superalgebras $F(4)$ and $G(3)$, respectively.
In the $F(4)$ case
\footnote{Recently Fradkin and Linetsky \cite{FrLi} has constructed
a quantum version of this algebra by using a non Lie
superalgebraic approach.}, we have eight supercurrents which belong
to the $spin(7)$ representation of the $B_{3}$.
The $G_{2}$ ESA associated with $G(3)$ has seven supercurrents
which belong to the 7-dimensional fundamental representation
of $G_{2}$.
The constructions of these algebras are straightforward
by using the matrix (pseudo)representations \cite{DeNi}.
{}From the viewpoint of Lie superalgebra, these complete the
classification of extended superconformal algebras.
It seems quite interesting to investigate the geometrical meaning of
these algebras, because $spin(7)$ and $G_{2}$ appear in the context of
the classification of restricted holonomy of non-symmetric
spaces \cite{Ber}.

It is possible to extend the present analysis to the case
of Lie superalgebras $sl(M|N)$ or $osp(M|2N)$, which include an
even subalgebra $sl(N)$ or $sp(2N)$.
In the former case we get a $W sl(N)$ minimal model coupled to
$gl(M)$ Wess-Zumino-Novikov-Witten (WZNW) model with $2M$ fermionic
currents with a conformal weight $(N+1)/2$.
In the latter case we get a $C_{N}$ Toda field theory
coupled to the $SO(M)$ WZNW model with N fermionic current with
weight $(N+1)/2$.

In a forthcoming paper we shall discuss the quantum hamiltonian
reduction and free field representation of extended superconformal
algebras.
A Feigin-Fuchs type representation enables us to analyse the
degenerate representation of the algebra and the computation of
correlation functions using screening operators.

We would like to thank Jens Lyng Petersen for useful discussions.

\newpage
\begin{center}
Table 1: classification of extended superconformal algebras

\begin{tabular}{lll}  \hline \hline
Lie superalgebra & even subalgebra & extended superconformal algebra \\
\hline
  $A(n|1)$ ($n\geq 2$)
& $A_{n}\oplus A_{1}\oplus u(1)$
& $N=2(n+1)$ $u(n+1)$ ESA \\
  $A(1|1)$
& $A_{1}\oplus A_{1}$
& $N=4$ $sl(2)$ ESA \\
  $A(1|0)=C(2)$
& $A_{1}\oplus u(1)$
& $N=2$ superconformal algebra \\
  $B(n|1)$ ($n\geq 1$)
& $B_{n}\oplus C_{1}$
& $N=2n+1$ $so(2n+1)$ ESA \\
  $D(n|1)$ ($n\geq 2$)
& $D_{n}\oplus C_{1}$
& $N=2n$ $so(2n)$ ESA \\
  $D(2|1;\A)$
& $A_{1}\oplus A_{1}\oplus A_{1}$
& $N=4$ $sl(2)\oplus sl(2)$ ESA \\
  $F(4)$
& $B_{3}\oplus A_{1}$
& $N=8$ $spin(7)$ ESA \\
  $G(3)$
& $G_{2}\oplus A_{1}$
& $N=7$ $G_{2}$ ESA \\
\hline \hline
\end{tabular}
\end{center}

\newpage
\begin{thebibliography}{99}
\bibitem{Za}
A.B.~Zamolodchikov, Theor.~Math.~Phys. {\bf 63} (1985) 1205.

\bibitem{DrSo}
V.G.~Drinfeld and V.V.~Sokolov, J. Sov. Math. {\bf 30} (1984) 1975.

\bibitem{Be}
A. A. Belavin, in \lq \lq Quantum String Theory"
Proceedings in Physics vol. {\bf 31} (Springer Verlag, Berlin ,1989);
M. Bershadsky and H. Ooguri, \CMP {\bf 126} (1989) 429;
J.~Balog, L.~Feh\'er, P.~Forg\'acs, L.~O'Raifeartaigh and A.~Wipf,
Ann. Phys. {\bf 203} (1990) 76.

\bibitem{Po}
A.M.~Polyakov, \IJMPA{5} (1990) 833.

\bibitem{Sa}
M.V.~Saveliev, \MPLA{5} (1990) 2223;
M.~Bershadsky, \CMP {\bf 139} (1991)71;
I.~Bakas and D.A.~Deprieux, \MPLA{6} (1991) 1561;
M.F.de~Groot, T.J.~Hollowood and L.J.~Miramontes,
preprint IASSNS-HEP-91/19.

\bibitem{FeOrRuTsWi}
L.~Feh\'er, L.~O'Raifeartaigh, P.~Ruelle and I.~Tsutsui and A.~Wipf,
Ann. Phys. {\bf 213} (1992) 1; preprint DIAS-STP-91-29;
P.~Bowcock and G.M.T.~Watts, preprint EFI 91-63.

\bibitem{BaTjDr}
F.A.~Bais, T.Tjin and P.~van Driel, \NPB{357} (1991) 632;
T.~Tjin and P.van~Driel, preprint ITFA-91-04;
J.~Fuchs, \PLB{262} (1991) 249;
L.~Romans, \NPB{B357} (1991) 549.

\bibitem{EvHo}
J.~Evans and T.~Hollowood, \NPB{352} (1991) 723;
S.~Komata, K.~Mohri and H.~Nohara, \NPB{359} (1991) 168;
T.~Inami and K.-I.~Izawa, \PLB{255} 521.

\bibitem{KnBe}
V.G.~Knizhnik, Theor.~Math.~Phys. {\bf 66} (1986) 68;
M.~Bershadsky, \PLB{174} (1986) 285.

\bibitem{Kup}
B.A.~Kupershmidt, \PL {\bf A109} (1985) (1985) 417;
T.G.~Khovanova, Theor. Math. Phys. {\bf 72} (1987) 899;
M. Bershadsky and H. Ooguri, Phys. Lett. {\bf B229} (1989) 374.

\bibitem{Mat}
P.~Mathieu, \PLB{218} (1989) 185.

\bibitem{Kac}
V.G.~Kac, Adv. Math. {\bf 26} (1977) 8;
Lect. Notes in Math. vol. {\bf 676}, p.597
(Springer-Verlag, Berlin, 1978).

\bibitem{ChKu}
M.~Chaichian and P.P.~Kulish, \PLB{183} (1987) 169.

\bibitem{GoSc}
P.~Goddard and A.~Schwimmer, \PLB{214} (1988) 209.

\bibitem{PeTa}
J.L.~Petersen and A.~Taormina, \NPB{331} (1990) 556;
{\bf B333} (1990) 833.

\bibitem{SeTrPr}
A.~Sevrin, W.~Troost and A.van~Proeyen, \PLB{208} (1988) 477.

\bibitem{FrLi}
E.S.~Fradkin and V.Ya.~Linetsky, ICTP preprint IC/91/348.

\bibitem{DeNi}
B.S.~DeWitt and P.van~Nieuwenhuizen,
J. Math. Phys. {\bf 23} (1982) 1953.

\bibitem{Ber}
M.~Berger, Bull. Soc. Math. France, {\bf 83} (1955) 279;
R.L.~Bryant, Lect. Notes in Math., vol. {\bf 1111}, p. 269
(Springer-Verlag, Berlin, 1985).

\end{thebibliography}
\end{document}